\documentclass{article}
\usepackage{amsmath,amssymb,capt-of,ifthen,calc,graphicx,bm,soul}
\usepackage[a4paper,top=2.5cm,bottom=2.5cm,right=2cm,left=2cm]{geometry}

\title{Multiscale lattice Boltzmann approach to modeling gas flows}

  \author{Jianping Meng$^1$, Yonghao Zhang$^1$ and Xiaowen Shan$^2$ \\
\it \small $^1$Department of Mechanical Engineering, University of
    Strathclyde, Glasgow G1 1XJ, UK \\
\it \small $^2$ Exa Corporation, 55 Network Drive, Burlington, Massachusetts 01803, USA
}
 
\begin{document}
\maketitle
  \begin{abstract} 
    For multiscale gas flows, kinetic-continuum hybrid method is usually
    used to balance the computational accuracy and efficiency. However,
    the kinetic-continuum coupling is not straightforward since the
    coupled methods are based on different theoretical
    frameworks. In particular, it is not easy to recover the
    non-equilibrium information required by the kinetic method which is lost by the
    continuum model at the coupling interface.  Therefore, we present a multiscale lattice Boltzmann
    (LB) method which deploys high-order LB models in highly rarefied flow
    regions and low-order ones in less rarefied regions. Since this
    multiscale approach is based on the same theoretical framework, the
    coupling precess becomes simple. The non-equilibrium
    information will not be lost at the interface as low-order LB models
    can also retain this information. The simulation results confirm that the
    present method can achieve model accuracy with reduced
    computational cost.
 \end{abstract}

\section{Introduction}

Many engineering problems involve multi-scale gas flows, e.g. gas flows in
micro/nano-fluidic devices \cite{citeulike:2623513}.  Since the flow regions can be
highly rarefied (non-equilibrium), the conventional continuum
theory becomes inappropriate. The rarefaction order of gas flows can be
classified by the non-dimensional Knudsen number, Kn, defined as the ratio
of the mean free path and the device characteristic length scale. The Navier
Stokes equations with no-velocity-slip wall boundary condition are only
appropriate in the hydrodynamic regime where $Kn<0.001$. In the slip flow
regime ($0.001<Kn<0.1$), or the transition flow regime ($0.1<Kn<10$), it
is necessary to use kinetic methods, e.g. direct simulation Monte Carlo
method (DSMC) method, to describe gas flows. Although the kinetic methods
including DSMC are able to simulate flow in the continuum or
near-continuum regimes, the computational cost is often very expensive
especially for low speed flows. Therefore, kinetic-continuum hybrid
methods are naturally employed to deal with mixed flow regimes in typical
microfluidic devices operating with a range of Knudsen numbers in
different parts, i.e., numerically efficient continuum approach will be
employed for the continuum regimes and kinetic approach for the rarefied
regimes(see Refs.
\cite{2009RSPSA.465.1581L,Tiwari2009,Sun2004,Schwartzentruber2007,Schwartzentruber2006,Wijesinghe,Tallec1997,Bourgat1996,PhysRevE.52.R5792,Burt2009,Hadjiconstantinou2005,wang:91}
and references therein). The
two models are coupled by exchanging information over the \lq \lq
hand-shaking\rq\rq region or across an interface.

However, the kinetic-continuum coupling is not straightforward since two
types of methods are based on different theoretical frameworks. While
information transferring from the kinetic model to the continuum model is
usually a well-defined process, the reverse process is more problematic
\cite{Hadjiconstantinou2005}. It is difficult to recover non-equilibrium
information lost by the continuum solvers which is required by the kinetic
method. Although the kinetic model can provide necessary information for
the continuum model, it can be computationally expensive \cite{Sun2004}. The statistical noise associated with the particle methods may also
affect the accuracy and stability of the hybrid solver \cite{Hadjiconstantinou2005}.
To resolve these problems, we introduce a multiscale lattice Boltzmann (LB) method to use LB models which are based on a same theoretical framework.

The LB method has been proven to be able to simulate hydrodynamic flows with only minimal number of discrete velocities
(e.g., nine discrete velocities for a two-dimensional
problem) \cite{Qian1995,1998anrfm..30..329c,1992phr...222..145b,Aidun2010,RevModPhys.74.1203}. For
continuum problems, its applicability is ensured by the Chapman-Enskog
expansion. Due to its kinetic nature, the LB model has distinct advantages
over other continuum computational methods, including easy implementation
of multi-physical mechanisms and the boundary conditions for fluid/wall interactions.

The LB method may also offer a flexible framework for rarefied flows, which has recently been demonstrated extensively
\cite{PhysRevE.76.025701,zhang:047702,Toschi2005,sbragaglia:093602,Sbragaglia2006,tang:046701,zhang:046704,2006JFM...550..413S,Ansumali2007,
  Kim20088655, yudistiawan:016705,guo:074903, Aidun2010}. It was shown
that the key to capturing the rarefaction effects is to choose appropriate
discrete velocity sets. Generally, a high-order LB model with larger discrete
velocity set describes non-equilibrium effects better
\cite{Ansumali2007,Kim20088655,yudistiawan:016705,Meng2009}. In particular,
high-order models with modest discrete velocity sets can already accurately capture non-equilibrium effects in rarefied flows
over a range of Knudsen numbers \cite{Ansumali2007,Kim20088655,yudistiawan:016705,Meng2009}. 

Since the LB method offers a solution for simulating gas flows ranging from continuum to
rarefied, we can introduce a multiscale method to couple models based on the
same LB framework. This can be accomplished by employing higher-order LB
models for non-equilibrium flow regions and lower-order LB models for
hydrodynamic flow regions. Since the coupled LB models only differ in the chosen discrete velocities
without loss of kinetic information at the coupling interface, the
information exchange process can be simplified. In particular, non-equilibrium information can be retained in
lower-order LB models, which resolves an intrinsic obstacle associated with
kinetic-continuum hybrid methods. Meanwhile, the LB model can still reserve some advantages of
particle method while eliminating the statistical noise.

\section{Lattice Boltzmann method}

\subsection{Lattice Boltzmann equation}

Historically, the LB method was developed from the lattice gas cellular
automata. The purpose was to mimic the Navier-Stokes dynamics. However,
it was revealed that its applicability should not
be limited to the hydrodynamic level
\cite{2006JFM...550..413S,PhysRevE.55.R6333,PhysRevE.56.6811,PhysRevLett.80.65,PhysRevE.58.6855,Ansumali2003,Chikatamarla2006,Meng2009,Ansumali2007}. There
are different theoretical frameworks for LB models e.g., entropic LB models
\cite{PhysRevE.79.046701,Karlin1999, PhysRevE.62.7999,
  Ansumali2003,Chikatamarla2006}. Here, to demonstrate the coupling
scheme, we will adopt commonly-used LB models based on the Hermite expansion detailed in
Refs.\cite{2006JFM...550..413S,PhysRevE.55.R6333,PhysRevE.56.6811,PhysRevLett.80.65,PhysRevE.58.6855,PhysRevE.81.036702}. However,
the proposed coupling approach can be equally applied to different LB models.

The original BGK equation is given as:
\begin{equation}
  \label{dbgk}
  \frac{\partial f}{\partial t} + \bm \xi \cdot \nabla f + \bm g \cdot \nabla_\xi f=
  -\frac{p}{\mu}\left(f-f^{eq} \right),
\end{equation}
where $f$ denotes the distribution function, $\bm \xi$ the phase velocity,
$p$ the pressure, $\bm g$ the body force and $\mu$ the gas
viscosity. Using the well-known Chapman-Enskog expansion, the collision
frequency can be represented by the ratio of pressure and gas viscosity,
which is convenient to obtain the Knudsen number definition consistent
with that of hydrodynamic models. Without losing generality, One can
define the following non-dimensional variables:
\begin{eqnarray}
  & \displaystyle \hat{\bm r}=\frac{\bm r}{l_0}, \hat{\bm u}=\frac{\bm u}{\sqrt{R T_0}},\hat{t}=\frac{\sqrt{R T_0} t}{l_0},
  \nonumber\\
  & \displaystyle \hat{\bm g}=\frac{l_0 \bm g}{ R T_0}, \hat{\bm \xi}= \frac{\bm \xi}{\sqrt{R T_0}}, \hat{T}=\frac{T}{T_0}, &
  \label{nondime}
\end{eqnarray} 
where $\bm u$ is the macroscopic velocity, $R$ the gas constant, $T$ the
gas temperature, $T_0$ the reference temperature, $\bm r$ the spatial
position and $l_0$ the characteristic length of the flow system. The
symbol {\it hat}, which denotes a dimensionless value, will hereinafter be
omitted. The Knudsen number can be defined by using macroscopic properties
as below:
\begin{equation}
  \label{KN}
  Kn=\frac{\mu \sqrt{RT_0}}{p l_0}.
\end{equation}
Based on these non-dimensional variables, the non-dimensional form of the
BGK equation becomes
\begin{equation}
  \label{bgk}
  \frac{\partial f}{\partial t} + \bm \xi \cdot \nabla f + \bm g \cdot \nabla_\xi f=
  -\frac{1}{Kn}\left(f-f^{eq} \right),
\end{equation}
where the Maxwell distribution in $D$-dimensional Cartesian coordinates
can be written as
\begin{equation}
  f^{eq}=\frac{\rho}{(2\pi T)^{D/2}} \exp\left[\frac{-(\bm \xi-\bm u)^2}{2 T}\right].
\end{equation}

For solving Eq.(\ref{bgk}), the velocity space can be firstly discretized
by projecting the distribution function onto a functional space spanned by
the orthogonal Hermite basis\cite{2006JFM...550..413S,12197}:
\begin{equation}
  \label{approxf}
  f(\bm r,\bm \xi,t) \approx f^{N}(\bm r,\bm \xi,t)=\omega(\bm \xi) \sum^N_{n=0}\frac{1}{n!}\bm a^{(n)}(\bm r,t) \bm \chi^{(n)}(\bm \xi),
\end{equation}
where $\bm \chi^{(n)}$ is the $n$th order Hermite polynomial, and
$\omega(\bm \xi)$ is the weight function, which are given by
\begin{equation}
  \label{hermipol}
  \bm \chi^{(n)}(\bm \xi)=\frac{(-1)^n}{\omega{(\bm \xi)}}\nabla^n \omega(\bm \xi),
\end{equation}
\begin{equation}
  \omega(\bm \xi)=\frac{1}{(2 \pi)^{D/2}} \text{e}^{-\xi^2/2}.
\end{equation}
The coefficients $\bm a^{(n)}$ are
\begin{eqnarray}
  \label{an}  
  \bm a^{(n)} & = &\int f \bm \chi^{(n)} d\bm \xi \approx \int f^{(N)} \bm
  \chi^{(n)} d\bm \xi \\
  & = &\sum^d_{\alpha=1} \frac{w_\alpha}{\omega(\bm \xi_\alpha)}f^{(N)}(\bm r, \bm
  \xi_\alpha,t)\bm \chi^{(n)}(\bm \xi_\alpha).  \nonumber 
\end{eqnarray}
The equilibrium distribution should also be expanded as\cite{2006JFM...550..413S}
\begin{equation}
  \label{approxfeq}
  f^{eq} \approx \omega(\bm \xi) \sum^N_{n=0}\frac{1}{n!}\bm a_{eq}^{(n)} \bm \chi^{(n)}(\bm \xi),
\end{equation}
where the coefficient $a_{eq}^{(n)}$ for the equilibrium distribution is
\begin{equation}
  \bm a_{eq}^{(n)}=\int f^{eq}\bm \chi^{(n)} d\bm \xi. 
\end{equation}
$w_\alpha$ and $\bm \xi_\alpha$, $a=1,\cdots,d$, are the weights and
abscissae of a Gauss-Hermite quadrature of degree $\geq 2N$
respectively. Therefore, the Maxwell distribution is approximated by up to
$N$ Hermite polynomials. The body force term $F(\bm r,\bm \xi,t) = \bm g
\cdot \nabla_\xi f$ can also be approximated
as\cite{2006JFM...550..413S,PhysRevE.58.6855}

\begin{equation}
  F(\bm r,\bm \xi,t) =\omega \sum^N_{n=1}\frac{1}{(n-1)!}\bm g \bm a^{(n-1)}\bm \chi^{(n)}.
\end{equation}

It was shown that Eq.(\ref{bgk}) with the first-order Hermite expansion
is sufficient to capture the rarefaction effects for isothermal and
incompressible flows\cite{Meng2009}. On the other hand, the second
order expansion has been proven to be able to model various Navier-Stokes
level problems \cite{Aidun2010,1998anrfm..30..329c}. Therefore, the second order
approximation of the equilibrium distribution and the body force will be
used hereinafter, as given below:
\begin{equation}
  \label{demoeq}
  f^{eq} \approx \omega(\bm \xi) \rho \{ 1+\bm \xi \cdot \bm u +
  \frac{1}{2}[(\bm \xi \cdot \bm u)^2-u^2+(T-1)(\xi^2-D)] \},
\end{equation}
\begin{equation}
  \label{demoforce}
  F(\bm r,\bm \xi,t)  \approx \omega(\bm \xi) \rho \{\bm g \cdot \bm \xi
  +(\bm g \cdot \bm \xi )(\bm u \cdot \bm \xi)-\bm g \cdot \bm u \},
\end{equation}
where $T$ should be unity for isothermal problems and $\rho$ is constant
for incompressible problems.

The discrete velocity set is revealed to be of upmost importance in
determining model accuracy for rarefaction effects \cite{Meng2009}. For the Navier-Stokes level
problems, several sets have been found to be applicable, e.g., the well
known D2Q9\footnote{We follow
the conventional terminology for the LB models as first introduced in
Ref.\cite{qian1992} dubbed as DnQm model i.e. n dimensional model with m
discrete velocities} model for two-dimensional flows. To capture higher-order
rarefaction effects, more discrete velocities are required. Some
modest discrete velocity sets were shown (e.g., D2Q16 and D2Q36) to be able to capture non-equilibrium effects for flows over a broad range of
Knudsen numbers\cite{Kim20088655,Meng2009,Ansumali2007,yudistiawan:016705}. Nevertheless, highly accurate discrete velocity set is
required for the flows with large Knudsen number. Therefore, coupling high-order and low-order LB models can save computational costs without sacrificing simulation accuracy for gas flows with mixed Knudsen numbers.

Discrete velocity sets can be obtained from several ways, see
Refs.\cite{2006JFM...550..413S,PhysRevE.81.036702,PhysRevE.79.046701,Chikatamarla2006}.
A direct method is utilizing the roots of Hermite polynomials\cite{2006JFM...550..413S}. In one-dimension, the discrete
velocities $\xi_\alpha$ are just the roots of Hermite polynomials, and its
corresponding weights are determined by:
\begin{equation}
  \label{weight}
  w_\alpha=\frac{n!}{[n \chi^{n-1}(\xi_\alpha)]^2}.
\end{equation} 
For higher dimensions, the discrete velocity set can be constructed by using the
``production'' formulae \cite{2006JFM...550..413S}. 

Once the discrete velocity set is chosen, Eq.(\ref{bgk}) can be discretized as
\begin{equation}
  \label{lbgk}
  \frac{\partial f_\alpha}{\partial t}+\bm \xi_\alpha \cdot \nabla f_\alpha =-\frac{1}{Kn}\left(f_\alpha-f_\alpha^{eq}\right)+g_\alpha,
\end{equation}
where $f_\alpha=\frac{w_\alpha f(\bm r,\bm \xi_\alpha,t)}{\omega(\bm
  \xi_\alpha)}$, $f_\alpha^{eq}=\frac{w_\alpha f^{eq}(\bm r,\bm
  \xi_\alpha,t)}{\omega(\bm \xi_\alpha)}$ and $g_\alpha=\frac{w_\alpha
  F(\bm r,\bm \xi_\alpha,t)}{\omega(\bm \xi_\alpha)}$.  Therefore, the LB
equation, i.e. Eq.(\ref{lbgk}), is now obtained by discretizing
Eq.(\ref{bgk}) in the velocity space.

\subsection{Coupling scheme}
\label{subc:coup}
The key to success of a coupling scheme is appropriate bi-directional
extraction and transfer of information at the interface or \lq\lq
hand-shaking \rq\rq region.  Since only LB models are used here, the extraction and transfer of information is in principle
seamless.  Lower order LB models, {\it in their applicable capacity}, can also retain non-equilibrium information which is required by the
higher-order models. For instance, the D2Q16 model can already perform well for a range of Knudsen
numbers \cite{Ansumali2007,Kim20088655,yudistiawan:016705,Meng2009}. The D2Q9
model, which has been used for hydrodynamic simulations, may
also capture some non-equilibrium effects \cite{Ansumali2007,yudistiawan:016705}. This is very different from the continuum methods in
kinetic-continuum hybrid models where non-equilibrium information is lost.

To correctly transfer information across the interface between two LB
models with different discrete velocities, the interface can
  be treated as a \lq\lq virtual boundary\rq\rq. Since a properly determined
interface should be located at smooth regimes where lower-order models
are valid, the relevant information can be obtained by using
extrapolation and interpolation techniques. Firstly, the related
macroscopic quantities can be calculated by interpolation, so that
the equilibrium part of \lq\lq boundary conditions\rq\rq is able to be
obtained by the Maxwell-Boltzmann distribution. For the non-equilibrium
part of information, recall that not only higher-order but also lower-order LB models can produce
accurate non-equilibrium information in the interface flow
region. Moreover, the information provided by two models should be completely same on the \lq\lq boundary\rq\rq. Therefore, the non-equilibrium part of information on
the \lq\lq boundary\rq\rq for the low-order and high-order LB models can
be obtained via extrapolating information on the grids adjacent to the \lq
\lq boundary\rq\rq. It is interesting to note that similar techniques
have been used to construct the no-slip boundary condition for continuum
problems \cite{guo:2007,Zhao-Li2002,chen:2527}.

\begin{figure}
  \begin{center}
    \includegraphics[width=8cm]{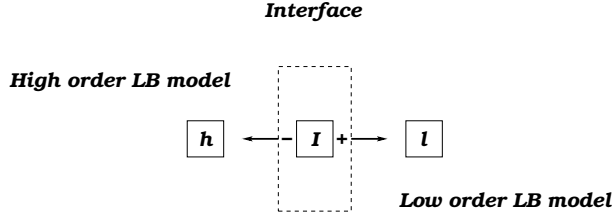}
    \caption{Schematic illustration of interface treatment where $I$ is the grid on the interface while $h$ and $l$ represent the adjacent grids at the computational domains for high-order and low-order LB models respectively.}
    \label{sketch}
  \end{center}
\end{figure}
To illustrate the scheme clearly, it is convenient to discuss in details for a
one-dimensional example. However, the same methodology can be generalized for
multi-dimensional problems. The distribution function can be decomposed into its equilibrium
($f_\alpha^{eq}(I,t)$) and non-equilibrium ($f_\alpha^{neq}(I,t)$)
parts, i.e.,
\begin{equation}
  f_\alpha(I,t) = f_\alpha^{eq}(I,t)+f_\alpha^{ne}(I,t),
\end{equation}
where the letter I denotes an interface grid (see Fig.\ref{sketch}). The
velocity direction needs to be further classified since different
discrete velocity sets are used across the interface. Hereinafter, the symbol $+$
denotes the discrete velocities (see
Fig.\ref{sketch}) pointing to the lower-order LB model side, $-$ to the
higher-order LB model side. Firstly, the macroscopic quantities related to
the equilibrium distribution can be obtained simply by the linear
interpolation, i.e.,
\begin{eqnarray}
  \rho_I=\frac{\rho_l+\rho_h}{2}, \\
  \bm u_I = \frac{\bm u_l+\bm u_h}{2},
\end{eqnarray} 
where $l$ and $h$ are the interface neighboring grids (see Fig.\ref{sketch}).
If necessary, the temperature can also be obtained similarly. With these
quantities, the equilibrium distribution can be written as

\begin{small}
  \begin{eqnarray}
    f_{\alpha+}^{eq}(I,t) \approx w_{\alpha+} \rho_I \left\{1+\bm \xi_{\alpha+} \cdot \bm u_I +
      \frac{1}{2}\left[(\bm \xi_{\alpha+} \cdot \bm u_I)^2-u_I^2+(T-1)(\xi_{\alpha+}^2-D)\right]
    \right\},   \\
    f_{\alpha -}^{eq}(I,t) \approx w_{\alpha -} \rho_I \left\{1+\bm \xi_{\alpha -} \cdot \bm u_I +
      \frac{1}{2}\left[(\bm \xi_{\alpha -} \cdot \bm u_I)^2-u_I^2+(T-1)(\xi_{\alpha -}^2-D)\right]
    \right\}.
  \end{eqnarray} 
\end{small}

Note, $\bm \xi_{\alpha -}$, $\bm \xi_{\alpha+}$ and $w_{\alpha-}$, $w_{\alpha+}$ belong to two different
discrete velocity sets. Based on the equilibrium distribution functions, the required
information can be transferred cross the interface. Meanwhile, a first
order extrapolation scheme is employed to supplement the information for
the non-equilibrium part, i.e.,
\begin{eqnarray}
  f_{\alpha+}^{neq}(I,t) = f_{\alpha+}(l,t)-f_{\alpha+}^{eq}(l,t), \\
  f_{\alpha -}^{neq}(I,t) = f_{\alpha -}(h,t)-f_{\alpha-}^{eq}(h,t).
\end{eqnarray} 

Therefore, the general process of the present multiscale LB simulation
starts from initialization to get all the necessary information, e.g., the
velocity field by utilizing either the lower-order or higher-order
model. The next step is to decompose the computational domain and determine the coupling
interface by choosing an appropriate switching criterion. The final step
is to implement the multiscale computation with lower-order models for
continuum or near-continuum regime, and higher-order models for more
rarefied regimes. Two models with different discrete velocity sets are coupled on the
interface as described above. The second and third steps are
repeated until the converged solutions are obtained.

The determination of interface, i.e., choosing an appropriate switching
criterion (also called \lq breakdown parameters\rq), is important to
any coupling/hybrid strategy. Several parameters have been proposed in literature,
e.g., the local Knudsen number based on the local spatial gradients of
hydrodynamic variables $ Kn_L=\frac{\lambda}{\phi}\left | \frac{d \phi}{d
    x} \right |$ ($\phi$ is the interested flow quantity, typically
density, temperature or pressure) \cite{boyd:210}, the \lq B\rq \
parameters $B=max\{\left|\tau_{ij}\right|, \left| q_i\right|\}$
($\tau_{ij}$ is stress and $q_i$, heat flux\cite{Garcia1998}). However,
different parameters may give significantly different values. So defining
an appropriate switching criterion remains an interesting problem in itself \cite{2009RSPSA.465.1581L,wang:91}. Here,
we do not intend to investigate the switching criterion and will
use what has been reported in the literature. In the next section, we will focus on numerical test
of the present multiscale LB method. 

\section{Numerical simulations and discussion}

\subsection{Numerical scheme}

\begin{figure}
  \begin{center}
    \includegraphics[width=8cm]{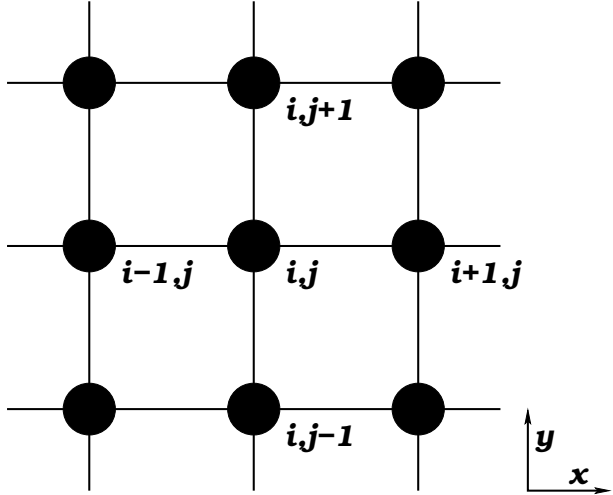}
    \caption{Schematic diagram of square lattices.}
    \label{figgrid}
  \end{center}
\end{figure}

To solve Eq.(\ref{lbgk}), various numerical schemes can be used. For
instance, if the first-order upwind finite difference scheme is chosen,
one can obtain the standard form of LB model, i.e., the stream-collision
mechanism. However, for some high-order LB models, the discrete velocity points
do not coincide with the lattice points. One has to choose a numerical
scheme to break the tie between the time step and the lattice spacing used
in the standard LB simulation\cite{Sbragaglia2010}. As some discontinuities may occur at wall
surface in the following simulations, we will employ the forward
Euler method for time discretization and the second-order total variation
diminishing (TVD) scheme for space discretization along the grid (see
Fig.\ref{figgrid}) for Eq.(\ref{lbgk})\cite{Kim20088655,Sofonea2009,PhysRevE.70.046702,695/BIBLTT}. According to the characteristics of problems, one can also
choose any other appropriate numerical method to solve Eq.(\ref{lbgk}).   

Let $f_{\alpha,i}^{n,j}$ denote the distribution function value $f_\alpha$
at the $n$th time step in the node ($x_i$, $y_j$) (see Fig.\ref{figgrid}),
the scheme can be written as
\begin{eqnarray}
  f_{\alpha,i}^{n+1,j} & = & f_{\alpha,i}^{n,j} - \frac{\xi_{\alpha
      x}\delta_t}{\delta_x}
  \left[\mathcal{F}_{\alpha,i+1/2}^{n,j}-\mathcal{F}_{\alpha,i-1/2}^{n,j}
  \right]\\
  & -&  \frac{\xi_{\alpha y}\delta_t}{\delta_y}
  \left[\mathcal{F}_{\alpha,i}^{n,j+1/2}-\mathcal{F}_{\alpha,i}^{n,j-1/2}
  \right] \nonumber \\
  &+&\frac{\delta_t}{Kn}(f^{eq,n,j}_{\alpha,i}-f_{\alpha,i}^{n,j})+g_\alpha\delta_t
  \nonumber,
\end{eqnarray} 
where $\delta_x$ and $\delta_y$ are the uniform grid spacing, and
$\delta_t$ is the time step, $\xi_{\alpha x}$ and $\xi_{\alpha y}$ denote
the phase velocity component at the $x$ and $y$ coordinates. The outgoing
and incoming fluxes in the node $(i,j)$ (see Fig.\ref{figgrid}) are
\begin{equation}
  \mathcal{F}_{\alpha,i+1/2}^{n,j}=f_{\alpha,i}^{n,j}+
  \frac{1}{2}\left(1-\frac{\xi_{\alpha x}\delta_t}{\delta_x} \right)\left[f_{\alpha,i+1}^{n,j}-f_{\alpha,i}^{n,j} \right]\Psi\left(\Theta^n_{\alpha,i}\right),  
\end{equation}
\begin{equation}
  \mathcal{F}_{\alpha,i-1/2}^{n,j}=\mathcal{F}_{\alpha,(i-1)+1/2}^{n,j},
\end{equation}
\begin{equation}
  \mathcal{F}_{\alpha,i}^{n,j+1/2}=f_{\alpha,i}^{n,j}+ \frac{1}{2}\left(1-\frac{\xi_{\alpha y}\delta_t}{\delta_y} \right)
  \left[f_{\alpha,i}^{n,j+1}-f_{\alpha,i}^{n,j} \right]\Psi\left(\Theta^{n,j}_{\alpha}\right),
\end{equation}
\begin{equation}
  \mathcal{F}_{\alpha,i}^{n,j-1/2}=\mathcal{F}_{\alpha,i}^{n,(j-1)+1/2},
\end{equation}
where
\begin{equation}
  \Theta_{\alpha,i}^{n}=\frac{f_{\alpha,i}^{n,j}-f_{\alpha,i-1}^{n,j}}{f_{\alpha,i+1}^{n,j}-f_{\alpha,i}^{n,j}},
\end{equation}
\begin{equation}
  \Theta_{\alpha}^{n,j}=\frac{f_{\alpha,i}^{n,j}-f_{\alpha,i}^{n,j-1}}{f_{\alpha,i}^{n,j+1}-f_{\alpha,i}^{n,j}},
\end{equation}
and the minmod flux limiter is
\begin{equation}
  \Psi\left(\Theta\right)= \max\left[0,\min(1,\Theta)\right].
\end{equation}

\subsection{Diffuse reflection boundary conditions}
\begin{figure}
  \begin{center}
    \includegraphics[width=8cm]{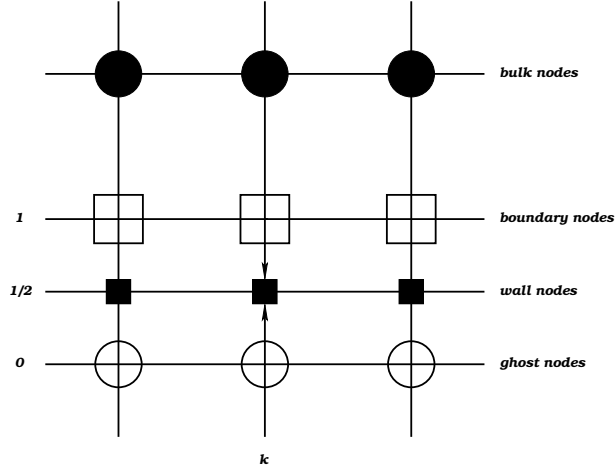}
    \caption{Schematic illustration of wall boundary treatment.}
    \label{figboundary}
  \end{center}
\end{figure}

Boundary treatment is of importance to correctly capture non-equilibrium
effects, e.g. flow characteristics in the Knudsen layer. The simple diffuse reflection model,
which was developed by Maxwell in 1879 \cite{Maxwell1879}, has been proved
to be sufficiently accurate for flows over a broad range of Knudsen
numbers.  The LB version of the Maxwellian model has also been developed
\cite{PhysRevE.66.026311}. Its specific numerical implementation on LB simulations
has been discussed in Refs.\cite{Sofonea2005639,Sofonea2009}.  In this
work, the Version 1 of boundary conditions in Ref.\cite{Sofonea2009} will
be employed.

For convenience, we assume
\begin{equation}
  S \approx w_\alpha \left\{1+\bm \xi_\alpha \cdot \bm u +
    \frac{1}{2}\left[(\bm \xi_\alpha \cdot \bm u)^2-u^2+(T-1)(\xi_\alpha^2-D)\right] \right\},
\end{equation}
i.e., $f^{eq}_\alpha = \rho S$.  As the discretization is
conducted along a Cartesian coordinate system (see Fig.\ref{figboundary}), the treatment of
wall boundary can be described as
\begin{equation}
  f^0_{\alpha,k}=\rho_{W,k} S(T_{W,k},\bm u_{W,k}) \mbox{\ \ \    } \bm \xi_\alpha \cdot \bm n >0 
\end{equation}
\begin{equation}
  \displaystyle
  \rho_{W,k}=\frac{\sum\limits_{(\bm \xi_{\alpha} \cdot \bm n)<0}  \left |\bm
    \xi_{\alpha} \cdot \bm n\right | f_{\alpha,k}^{1}}{\sum\limits_{(\bm
    \xi_{\alpha} \cdot \bm n)>0}  \left |\bm
    \xi_{\alpha} \cdot \bm n\right | S(T_{W,k},\bm u_{W,k}) },  
\end{equation}
where the subscript $W$ denotes the computational nodes at the wall, $\rho_{W,k}$ denotes the density on the wall nodes $k$ (see
Fig.\ref{figboundary}), $T_{W,k}$, the temperature, $\bm u_{W,k}$, the
velocity, $\bm n$, the unity normal vector to the wall. Here, the
distribution function in the ghost nodes are assumed to be identical to
those on the corresponding wall nodes.

\subsection{Kramers' problem}

The classic Kramers' problem is often used to assess model capability in
capturing the flow characteristics in the Knudsen layer (up to a few mean
free paths away from the wall). In this problem, a gas fills the
half-space ($y>0$) bounded by a plate at $y=0$. A constant shear rate is
applied along the plate at $y \rightarrow \infty$. With this special
setup, one can investigate the nonlinear Knudsen layer in detail. To
correctly predict this Knudsen layer, a kinetic method is
required. However, for the flow region far from the plate wall, a
continuum method is sufficient. Therefore, the problem is appropriate to
test the coupling approach described in Sec. \ref{subc:coup}.

\begin{figure}
  \centering
  \includegraphics[width=8cm,height=5cm]{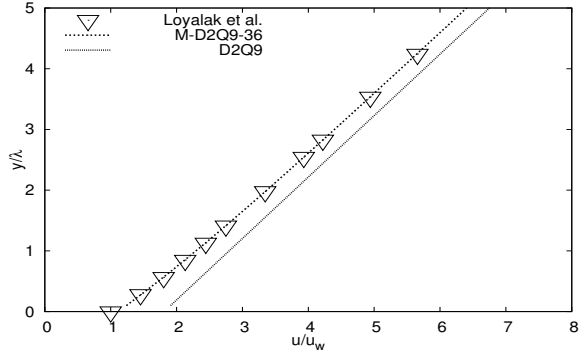}
  \caption{The velocity profile of Kramers' problem. The symbols are the
    data from Loyalka \textit{et al.} \cite{loyalka:1094}, where the BGK equation was directly solved.}
  \label{krame}
\end{figure}

In the simulations, the plate is fixed at $y=0$ and a constant shear rate
is applied at $y=200\lambda$($\lambda$ denotes the mean free path). The
Maxwellian diffuse reflection boundary condition is employed for the fixed
wall. The D2Q36 LB model is used for the region near the plate (up to
$10\lambda$ from the wall) and the D2Q9 model for the other
region\footnote{For convenience, the multiscale LB model will be named according
  to the rule M-D2Qi-j where i and j denote the discrete velocities of
  lower-order and higher-order model respectively.}. The
results in Fig.\ref{krame} show that the nonlinear velocity profile is
captured well by the multiscale LB method. It indicates that the coupling process
can effectively exchange information bi-directionally.
 
\subsection{Steady Couette flow}

\begin{figure}
  \centering
  \includegraphics[width=8cm,height=5cm]{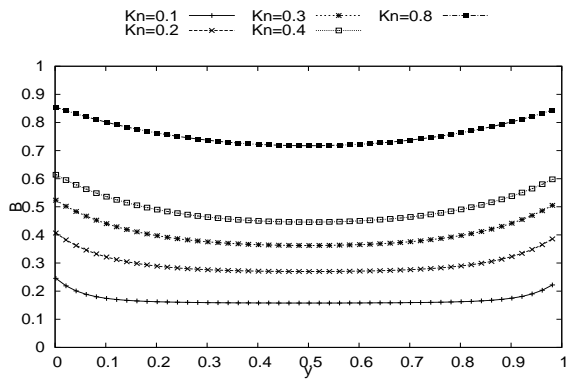}
  \caption{The \lq B\rq profile \cite{Garcia1998} for the Couette flows at
    different global Knudsen numbers.}
  \label{lk}
\end{figure}

With a simple geometrical configure, Couette flow represents many
realistic shear dominant applications, e.g. reader heads of a hard-disc
driver, micro turbines and gas bearings. Moreover, Couette flow is a
theoretically well defined problem. Therefore, it is generally used as a
benchmark problem. Particularly, its geometry is so simple that the
coupling interface can be determined easily. As has been shown by
Fig.\ref{lk}, the \lq B\rq profiles \cite{Garcia1998} indicate that the flow
regimes near the wall are highly rarefied and the discontinuities
occur at the wall. So we can use higher-order LB models in the near-wall
regions and lower-order models in the middle. Therefore, the Couette flow
can serve as a benchmark problem for testing the coupling strategy.

In the following simulations, the lower-order LB model will be employed
for $70\%$ of the computational region in the middle and the higher-order
model for the other regions adjacent to the walls. The upper and lower plates
are set to be moving oppositely with the same velocity magnitude, and the
diffuse boundary condition is used for gas/wall interactions.

\begin{figure}
  \centering
  \includegraphics[width=8cm,height=5cm]{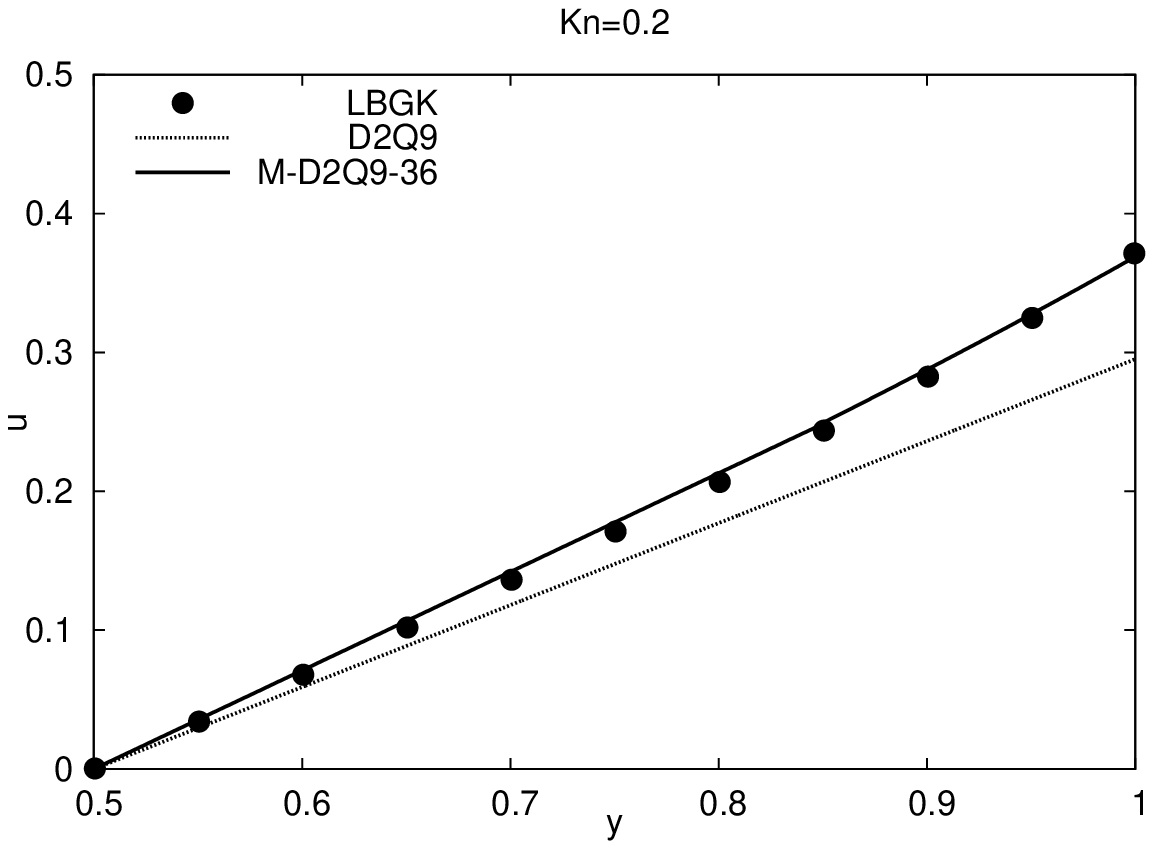}
  \includegraphics[width=8cm,height=5cm]{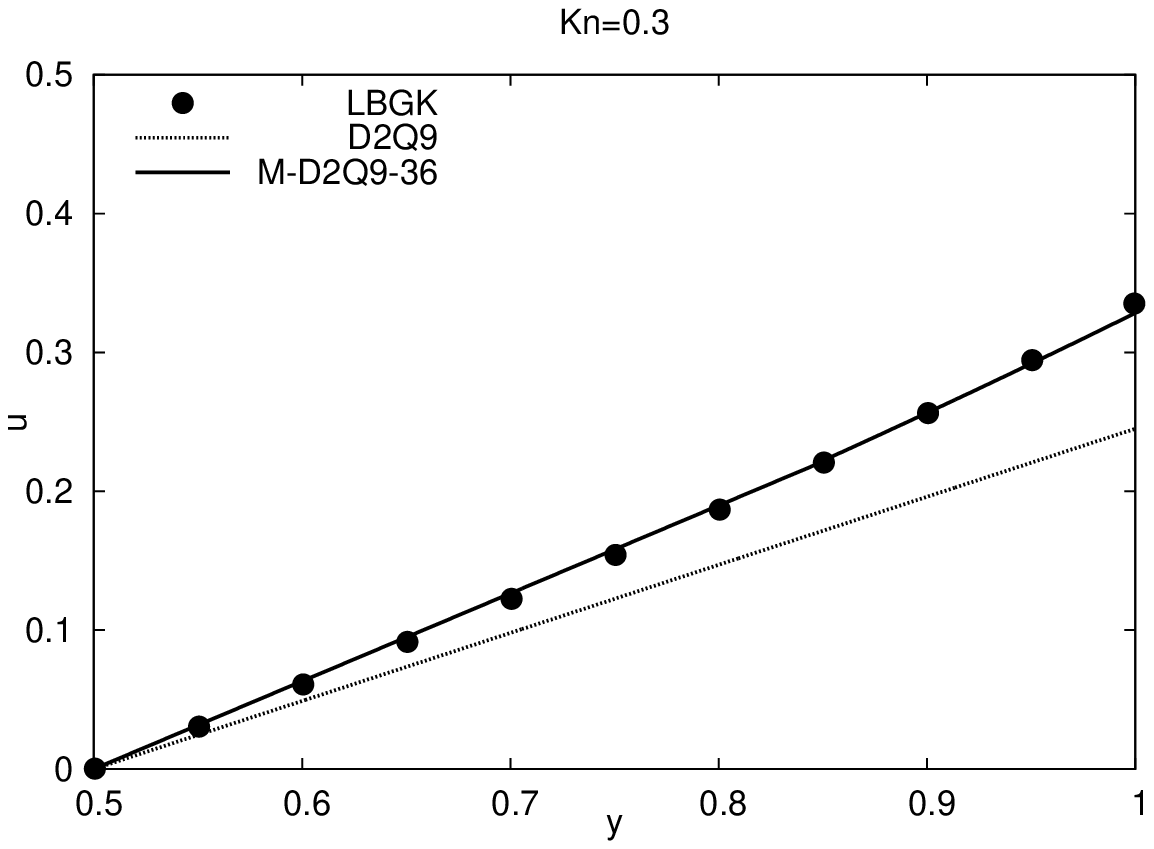}
  \includegraphics[width=8cm,height=5cm]{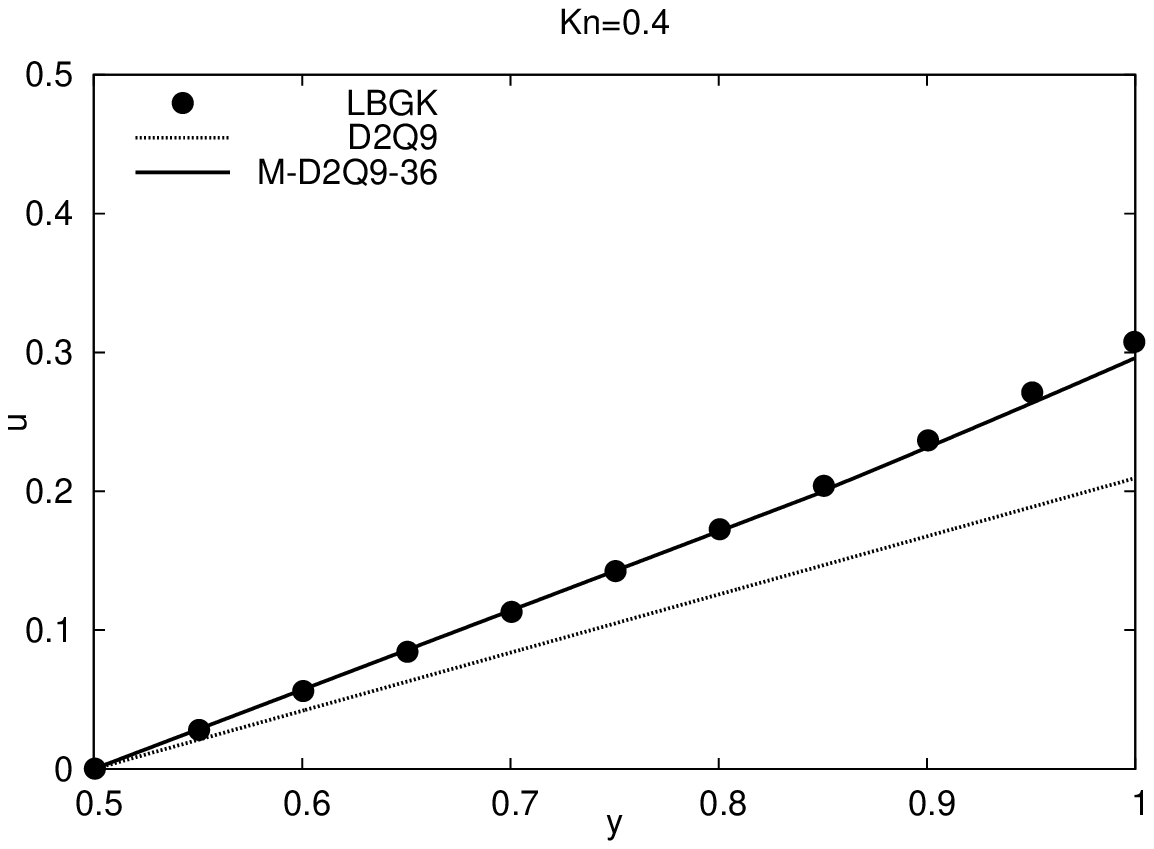}
  \includegraphics[width=8cm,height=5cm]{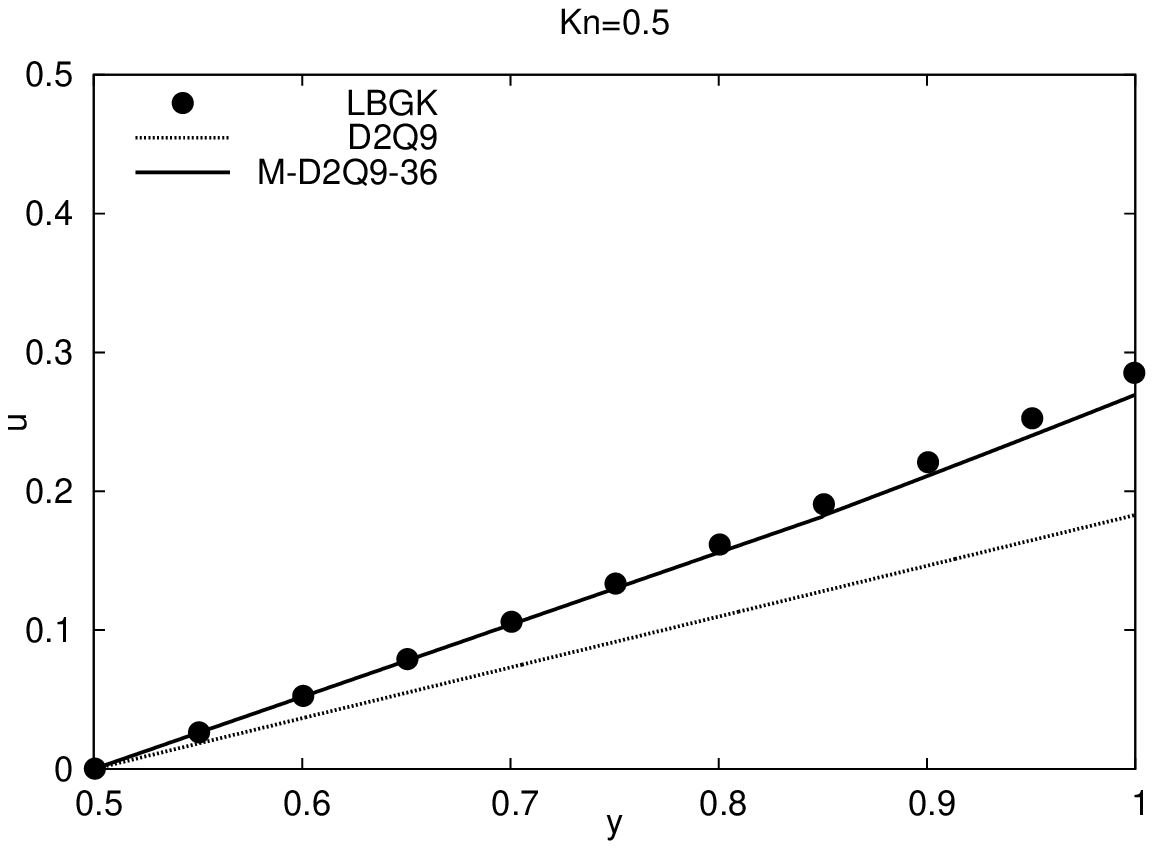}
  \caption{Nondimensional velocity profiles for planar Couette flows where
    the velocity is normalized by the velocity difference between the two
    plates.}
  \label{d0936}
\end{figure}

In Fig.\ref{d0936}, it is clear that the D2Q9 model is unable to describe
the Knudsen layer, which was also reported previously \cite{Ansumali2007,Kim20088655,yudistiawan:016705}, while the M-D2Q9-36 model can obtain
satisfactory results with the global Knudsen number up to $0.5$. When the
global Knudsen number is larger than $0.5$, the multiscale method starts to
deviate more from the linearized BGK (LBGK) results. This is not surprising since the
Knudsen layers overlap and rarefaction effect becomes important for the
whole flow domain (cf. Fig.\ref{lk}). Note, a typical Navier-Stokes and
DSMC hybrid model usually become problematic when the Knudsen number is
over $0.1$, e.g., see Fig.4 in Ref.\cite{Sun2004}. To some extent, this
indicates the advantage of coupling the kinetic-based LB models. 

\begin{figure}
  \centering
  \includegraphics[width=8cm,height=5cm]{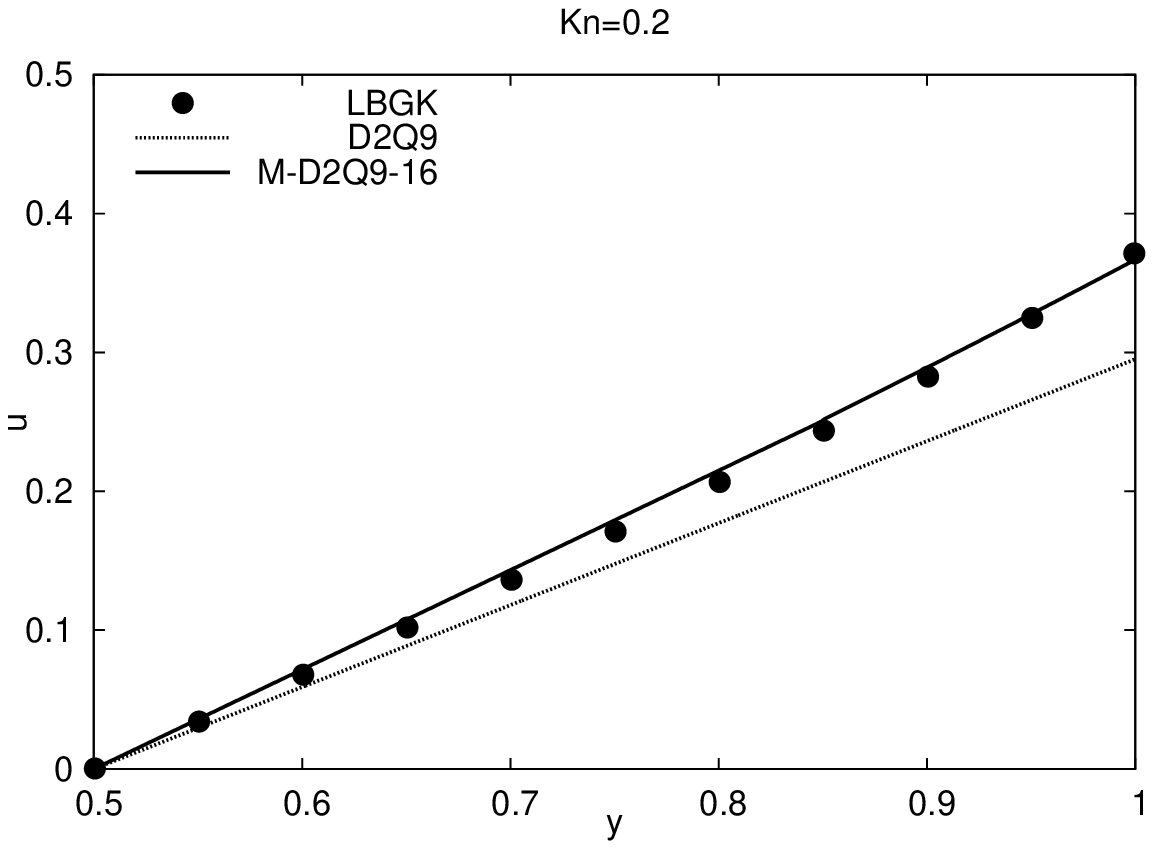}
  \includegraphics[width=8cm,height=5cm]{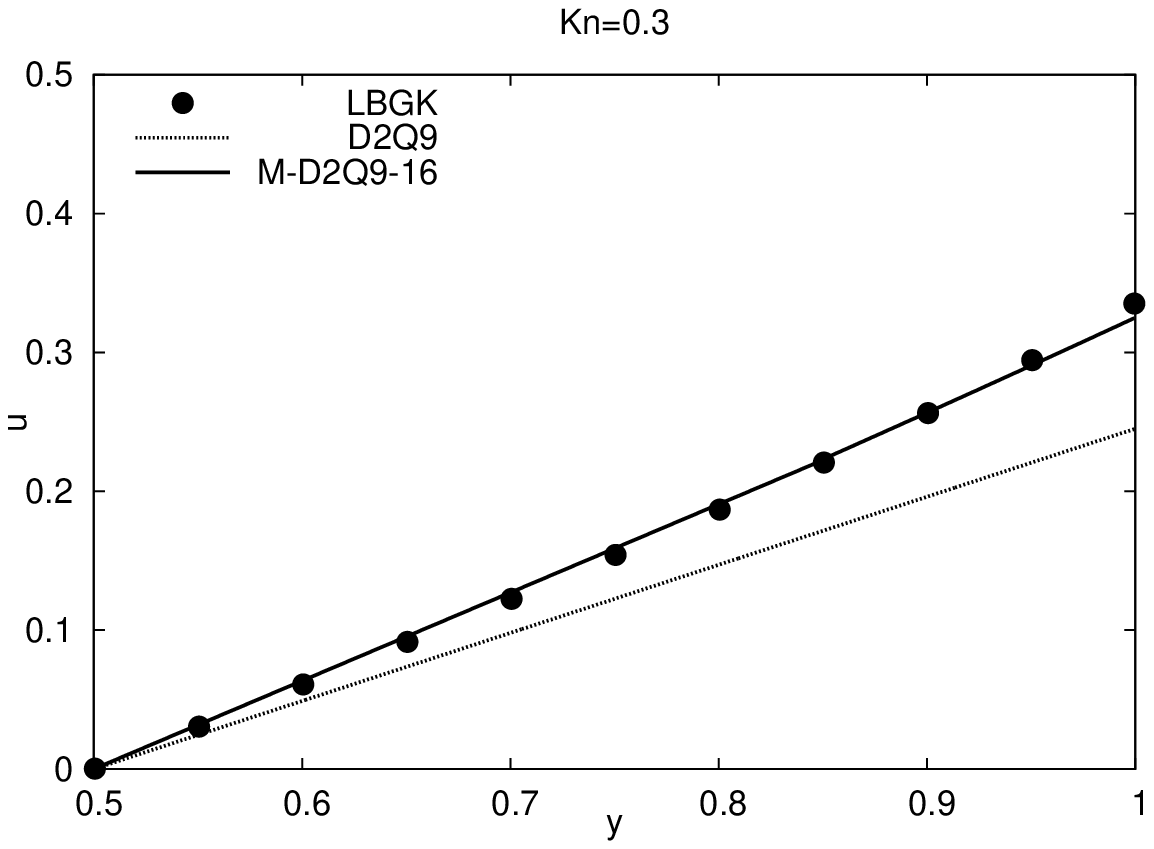}
  \includegraphics[width=8cm,height=5cm]{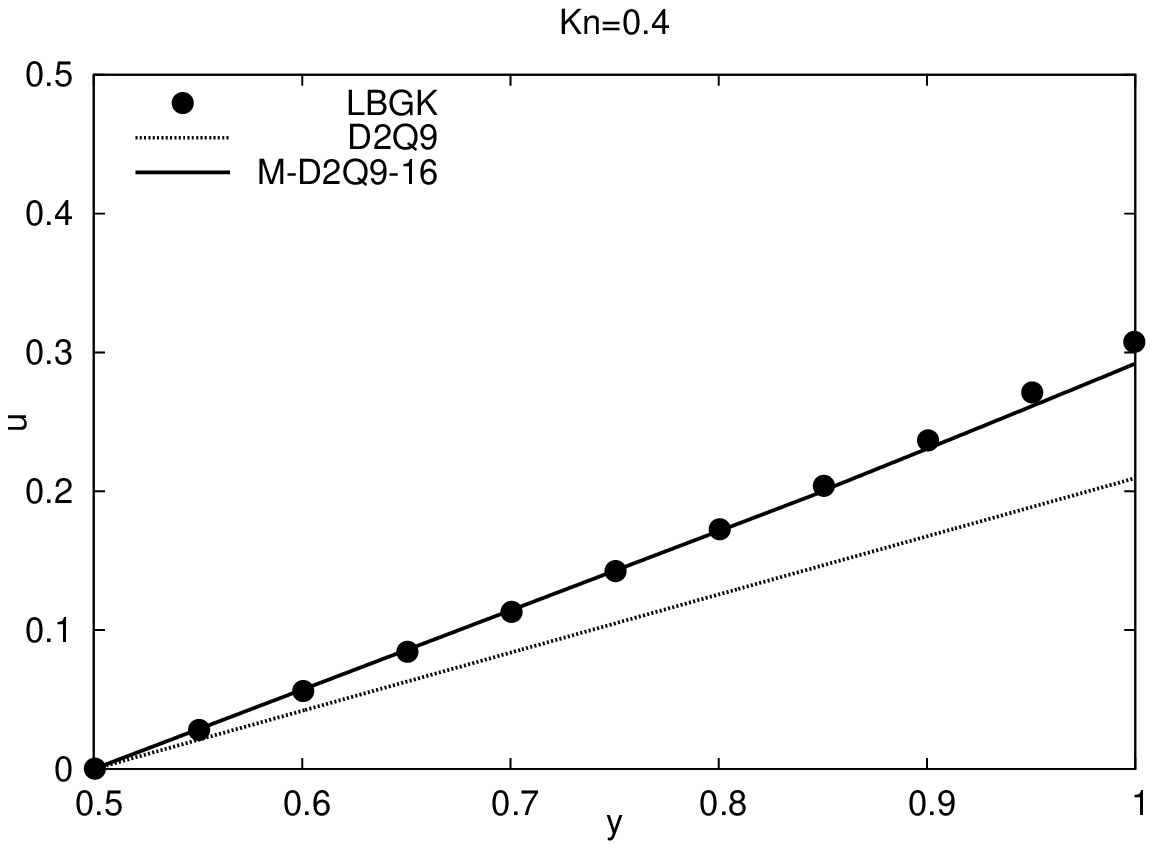}
  \includegraphics[width=8cm,height=5cm]{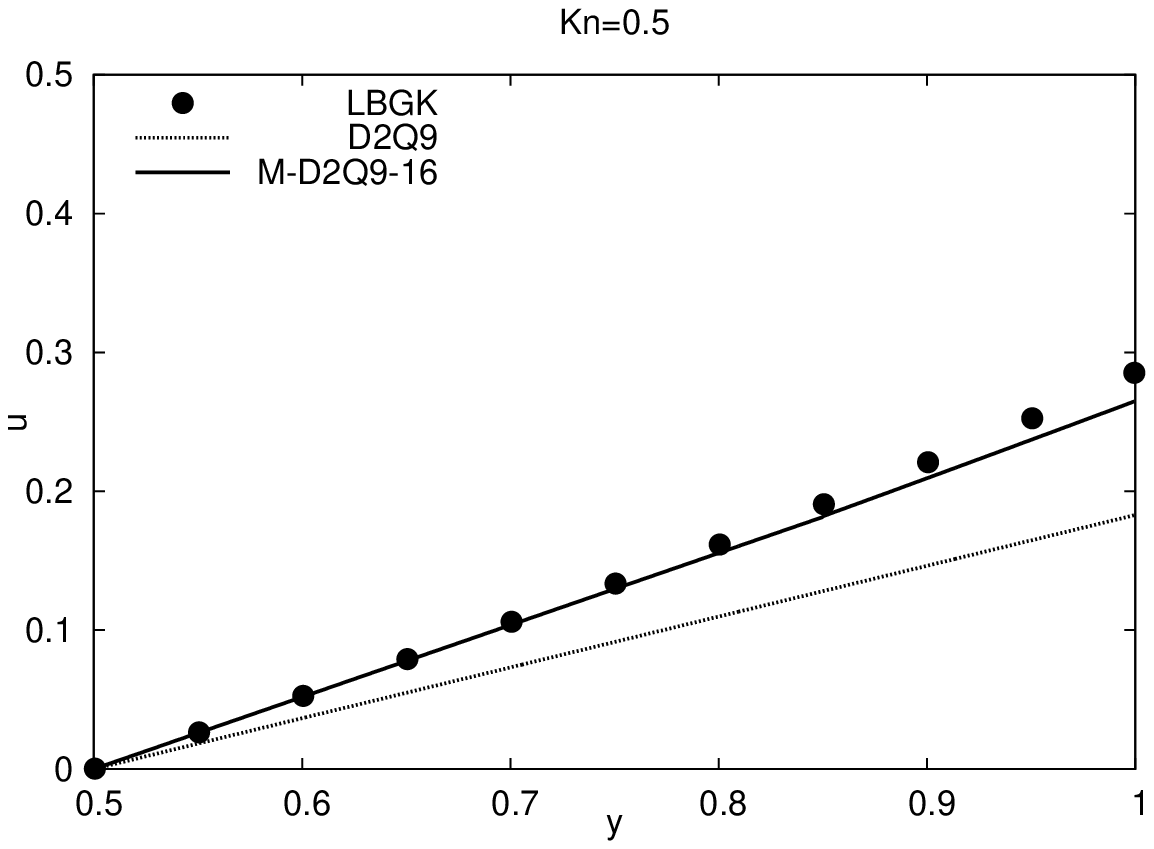}
  \caption{Nondimensional velocity profiles for the planar Couette flows where
    the velocity is normalized by the velocity difference between the two
    plates.}
  \label{d0916}
\end{figure}

As has been shown\cite{Meng2009,Kim20088655,Ansumali2007,yudistiawan:016705},
various higher-order LB models can satisfy different requirement on model accuracy in terms of capturing high-order rarefaction effects.  Therefore, it is possible to choose LB models with appropriate
discrete velocity sets according to the requirements of model accuracy and
computational cost. For instance, although the D2Q36 model is used for the
regions near the wall in the above simulations, the D2Q16 model may also be able to perform
well for Knudsen numbers up to 0.4, see Fig.\ref{d0916}. Therefore, there is some flexibility in choosing various-order LB
models for the present multiscale method.
    
\subsection{Oscillatory Couette flow}
The oscillatory Couette flows can mimic flows in many microfluidic
devices containing oscillating parts. Its setup consists of a stationary
plate at $y=l_0$ and a moving plate at $y=0$ which oscillates harmonically
in the lateral direction with velocity $u=u_w\sin(\omega t)$. This flow
can be characterized by the Stokes number
\begin{equation}
  \beta =\sqrt{\frac{\rho \omega L^2}{\mu}},
\end{equation}
which represents the balance between the unsteady and viscous
effects. Similar to the steady case, $70\%$ of the computational domain is
computed with the lower-order LB model. The results will be compared to
those of the variance-reduced (VR) particle simulations and the VR method is discussed in the Ref.\cite{PhysRevE.79.056711}.

\begin{figure}
  \centering
  \includegraphics[width=8cm,height=5cm]{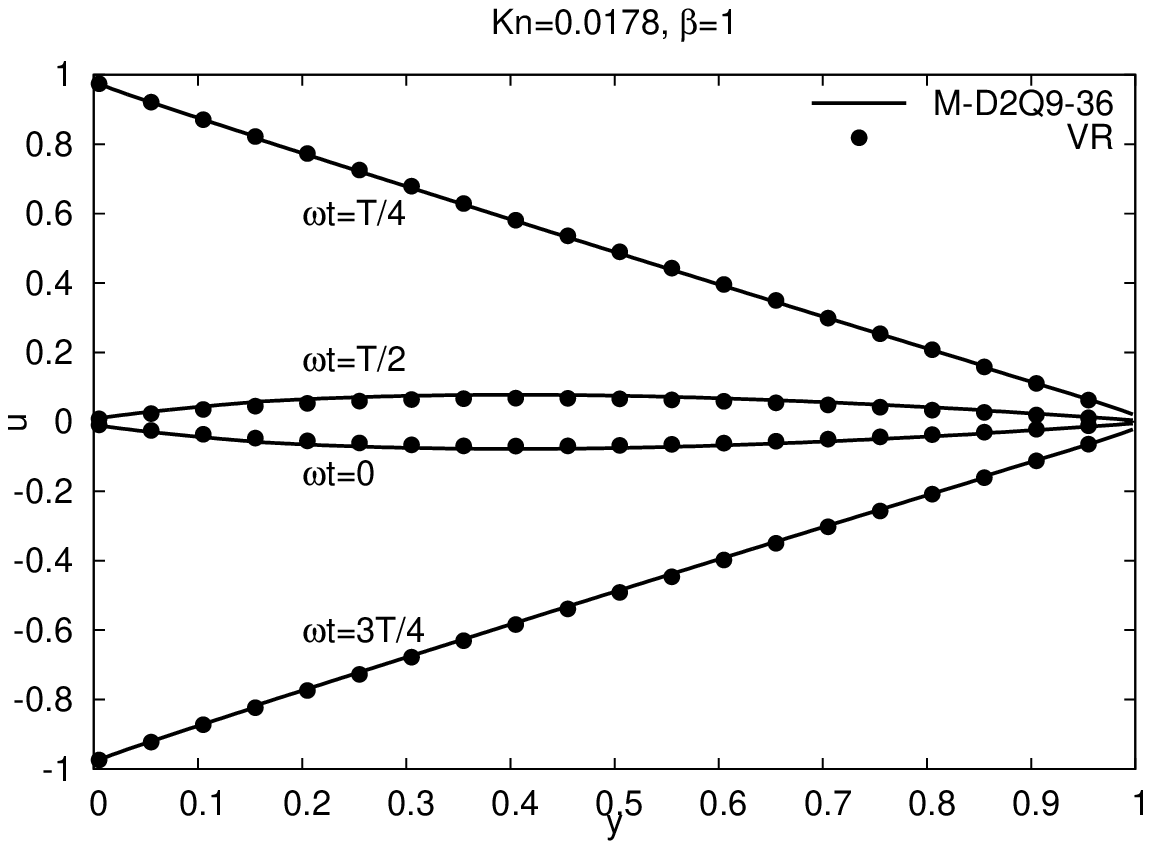}
  \includegraphics[width=8cm,height=5cm]{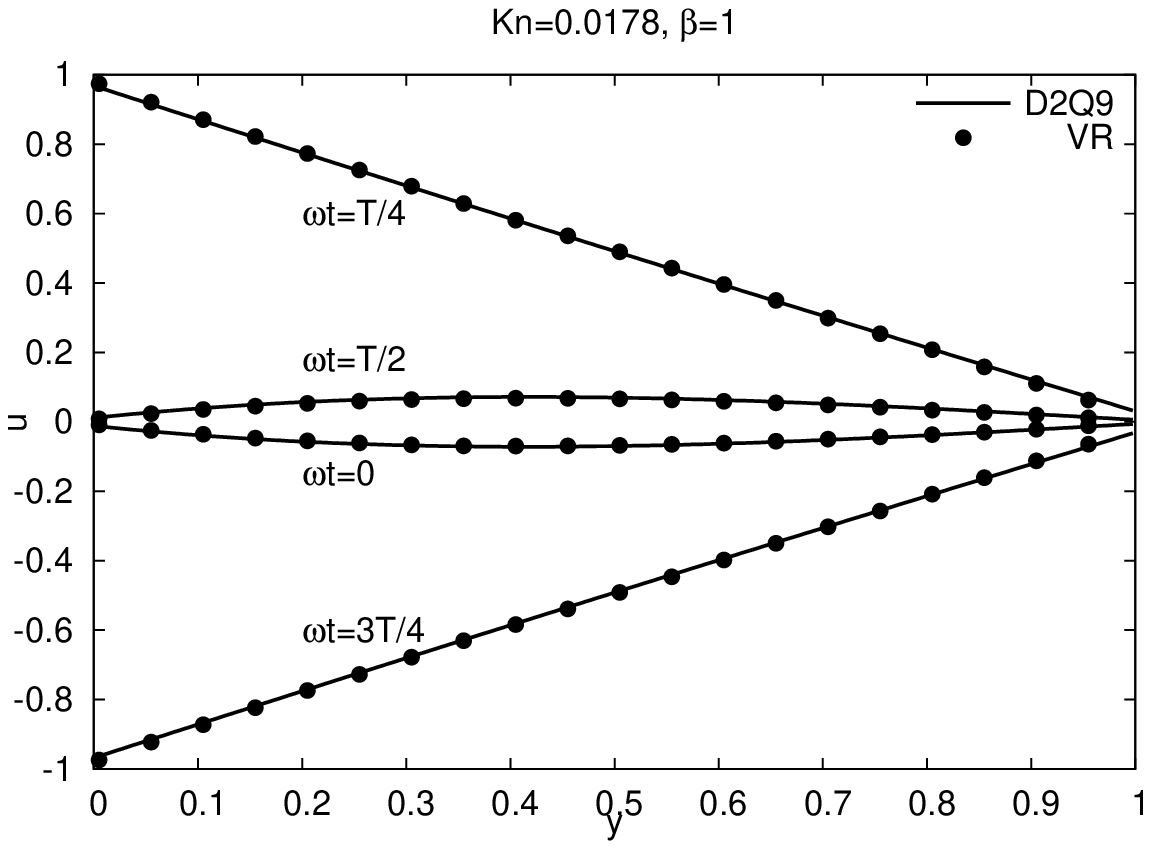} 
  \caption{Nondimensional dynamics velocity profiles for Oscillatory
    Couette flows where the velocity is normalized by the velocity
    amplitude of oscillating plate.}
   \label{osicu1}   
\end{figure}

\begin{figure}
  \centering
  \includegraphics[width=8cm,height=5cm]{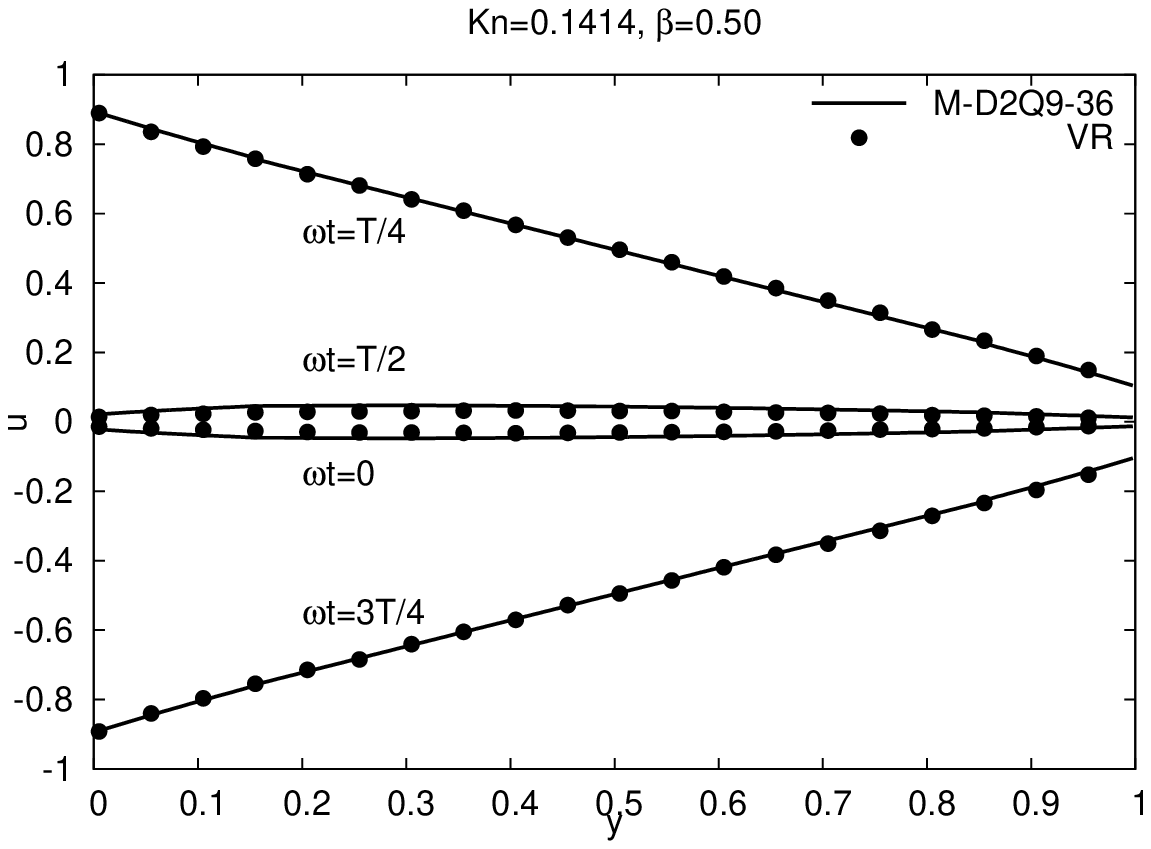}
  \includegraphics[width=8cm,height=5cm]{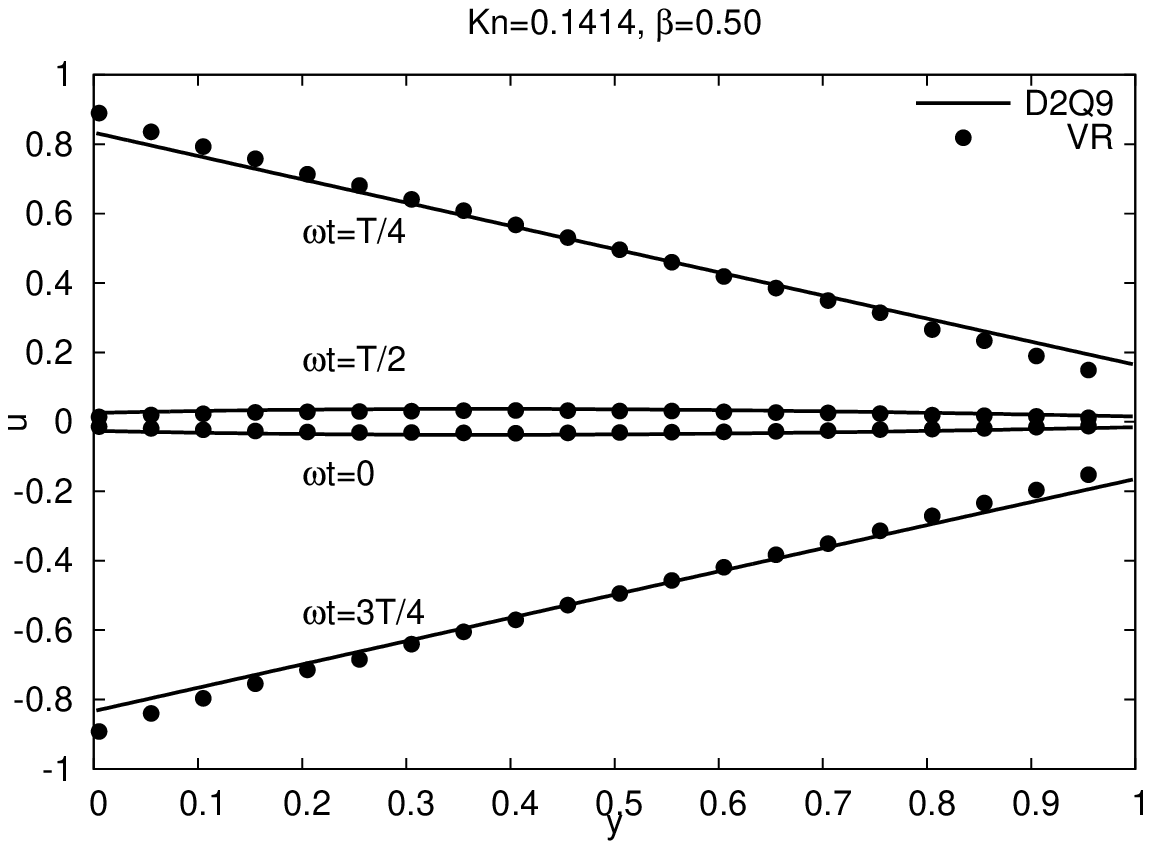} 
  \caption{Nondimensional dynamics velocity profiles for oscillatory
    Couette flows where the velocity is normalized by the velocity
    amplitude of oscillating plate.}
  \label{osicu2}
\end{figure}

\begin{figure}
  \centering
  \includegraphics[width=8cm,height=5cm]{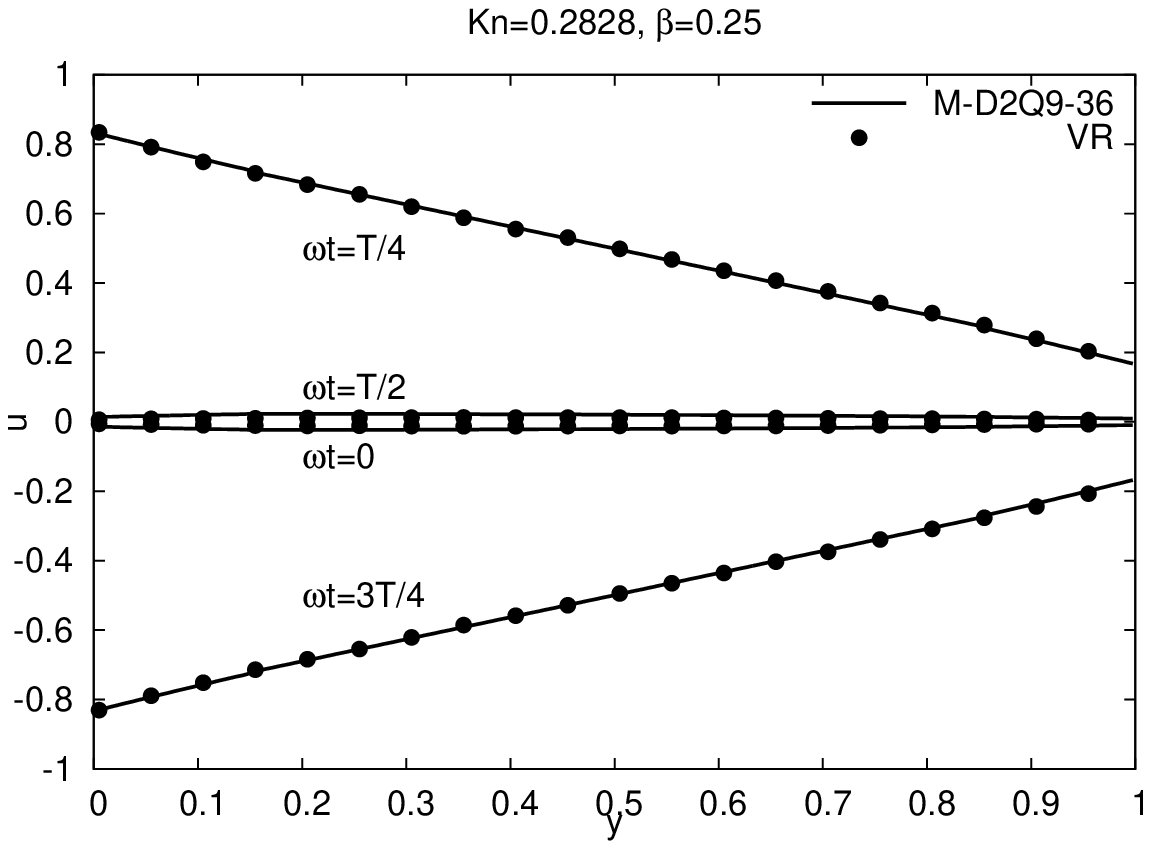}
  \includegraphics[width=8cm,height=5cm]{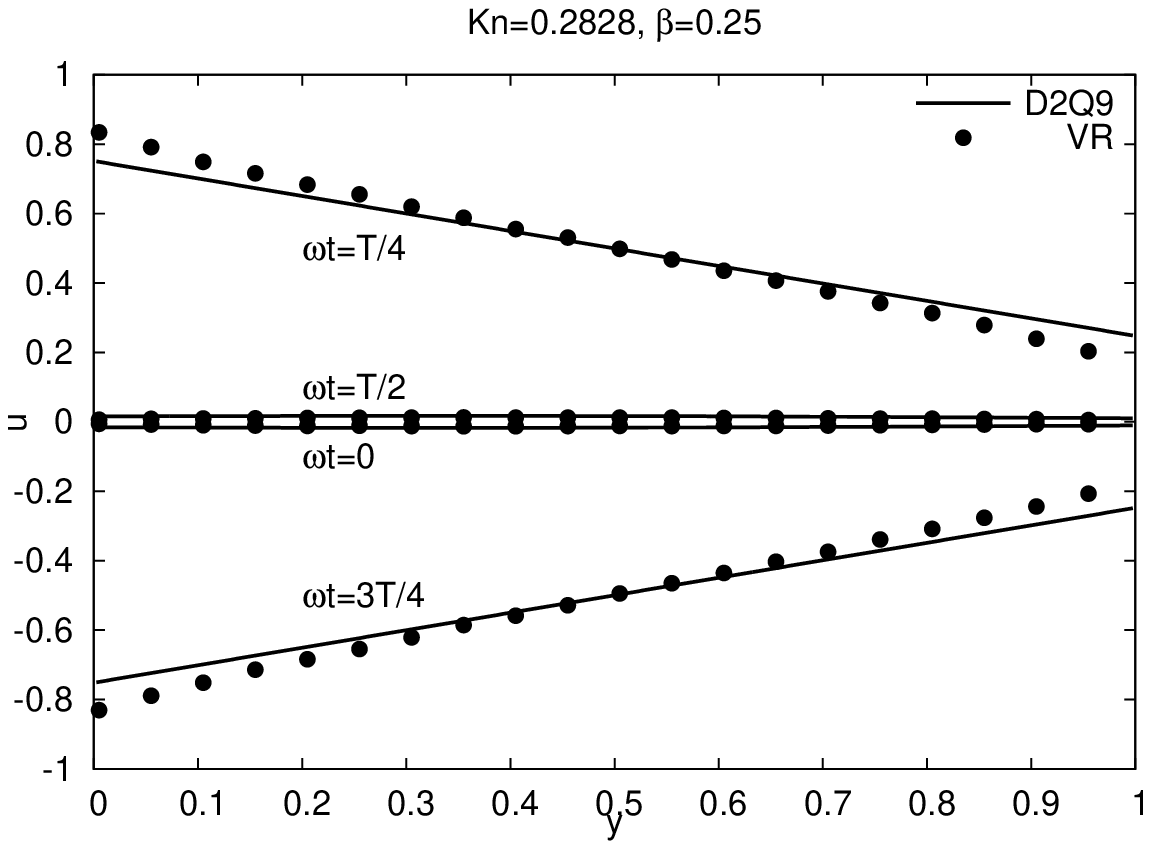} 
  \caption{Nondimensional dynamical velocity profiles for oscillatory
    Couette flows where the velocity is normalized by the velocity
    amplitude of the oscillating plate.}
  \label{osicu3}
\end{figure}

\begin{figure}
  \centering
  \includegraphics[width=8cm,height=5cm]{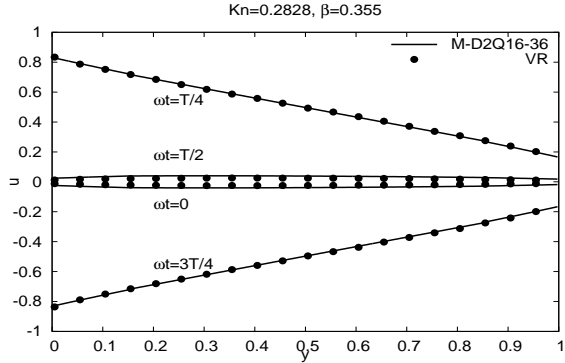}
   \caption{Nondimensional dynamics velocity profiles for oscillatory
    Couette flows where the velocity is normalized by the velocity
    amplitude of the oscillating plate.}
  \label{osicu4}
\end{figure}

Fig.\ref{osicu1} shows that both D2Q9 and M-D2Q9-36 models are valid in
the hydrodynamic regime when the Knudsen number is low (Kn=0.0178). When
the Knudsen number increases and the flows are in the transition regime,
Figs.\ref{osicu2} and \ref{osicu3} show that the D2Q9 model along becomes
inappropriate while the M-D2Q9-36 model still performs well. This
demonstrates that the present multiscale method can work well for the
flows with various degree of rarefaction. The simulation results of the M-D2Q16-36 model as presented in
Fig.\ref{osicu4} further indicates the flexibility in choosing various-order LB models.

It is also interesting to investigate the computational performance of the
multiscale method. We will test the computational performance of the D2Q9,
D2Q36 and multiscale models. For the M-D2Q9-36, $10\%$ of the flow region is
computed by the D2Q36 model and the rest is simulated by the D2Q9
model. The simulations are run on a four-core PC (Intel Core 2
QuadQ6600@2.4GHZ) without parallelization (i.e., only one core is utilized). The time
required for each computational step is 0.114 ms for the M-D2Q9-36
model, 0.250 ms for D2Q36 model and 0.097 ms for D2Q9 model
respectively. Therefore, the present multiscale approach can effectively reduce
the computational costs for mixed-Kn flows. Similar to other hybrid
methods, the performance of multiscale approach depends on how the computational
domain is divided and calculated by the lower and higher order LB
models. However, the LB framework can have some flexibility since various
discrete velocity sets can be chosen to satisfy the requirement on model
accuracy at the minimum computing cost. The details about how to choose
appropriate LB models can be found in many references e.g., Refs.\cite{2006JFM...550..413S,Ansumali2007,Kim20088655,yudistiawan:016705,Meng2009,PhysRevE.81.036702,PhysRevE.79.046701}

\section{Concluding remarks}

A multiscale LB method utilizing low-order and high-order LB models has been developed for gas flow simulation. As a hierarchy of LB models with various discrete velocity sets can be chosen, the multiscale method offers flexibility in designing
coupling strategy to strike appropriate balance between model accuracy and
computational efficiency. The present coupling process is simple by
using interpolation and extrapolation processes. Therefore, the difficulty associated with kinetic-continuum hybrid models which couple two different
methods becomes amenable. Furthermore, the present methodology can be
extended to develop other kinetic-kinetic hybrid models e.g. using
discrete velocity models.

\section{Acknowledgments}
The authors would like to thank Nicolas Hadjiconstantinou of MIT, USA, who
has performed the variance-reduced particle simulations and kindly provided us
the data. This work was financially supported by the Engineering and Physical Sciences Research
Council U.K. under Grants No. EP/ F028865/1. The research leading to these
results has received the funding from the European Community's Seventh
Framework Programme FP7/2007-2013 under grant agreement ITN GASMEMS No. 215504.\\
\\

\end{document}